\documentclass[prl,twocolumn,showpacs,amsmath,amssymb,superscriptaddress]{revtex4}
\usepackage{graphicx}
\usepackage{graphics}
\usepackage{dcolumn}
\usepackage{bm}

\begin{document}
\title{Universal Intrinsic Spin-Hall Effect}
\author{Jairo Sinova}
\affiliation{Department of Physics, Texas A\&M University, College Station, TX 77843-4242}
\affiliation{Department of Physics, University of Texas at Austin, Austin TX 78712-1081}
\author{Dimitrie Culcer}
\affiliation{Department of Physics, University of Texas at Austin, Austin TX 78712-1081}
\author{Q. Niu}
\affiliation{Department of Physics, University of Texas at Austin, Austin TX 78712-1081}
\author{N. A. Sinitsyn}
\affiliation{Department of Physics, Texas A\&M University, College Station, TX 77843-4242}
\author{T. Jungwirth}
\affiliation{Department of Physics, University of Texas at Austin, Austin TX 78712-1081}
\affiliation{ Institute of Physics  ASCR, Cukrovarnick\'a 10, 162 53
Praha 6, Czech Republic }
\author{A.H. MacDonald}
\affiliation{Department of Physics, University of Texas at Austin, Austin TX 78712-1081}
\date{\today}
\begin{abstract}
We describe a new effect in semiconductor spintronics that leads to 
dissipationless spin-currents in paramagnetic spin-orbit coupled systems. 
We argue that in a high mobility two-dimensional electron 
system with substantial Rashba spin-orbit coupling, a spin-current
that flows perpendicular to the charge current is intrinsic. 
In the usual case where both spin-orbit split bands
are occupied, the intrinsic spin-Hall conductivity has a universal 
value for zero quasiparticle spectral broadening.  
\end{abstract}

\pacs{72.10.-d, 72.15.Gd, 73.50.Jt}

\maketitle 

The science of devices whose operation is based in part on
manipulation of the electronic spin degree of freedom, spintronics,
has emerged \cite{wolf} as an active subfield of condensed matter physics
because of its potential impact on information technology and because 
of the challenging basic questions that it poses.
Many spintronic concepts involve ferromagnets, in which spins are 
easier to manipulate because they behave collectively. Spintronic 
magnetoresistive \cite{gmr,gmr2,gmr3,gmr4} sensors based on the properties of ferromagnetic
metals, for example, have reinvented the hard-disk industry over the past
several years.  Spintronics in semiconductors is richer scientifically
than spintronics in metals because doping, gating, and
heterojunction formation can be used to engineer key 
material properties, and because of the intimate relationship
in semiconductors between optical and transport properties. Practical
spintronics in semiconductors has appeared, however, to be contingent on either 
injection of spin-polarized carriers \cite{jonker,molenkamp,molenkamp2,spininject,
spininject2,spininject3,spininject4} from 
ferromagnetic metals combined with long spin lifetimes \cite{awschalom},
or on room-temperature semiconductor ferromagnetism \cite{ohno}. 
In this paper we explain a new effect \cite{murakami}
that might suggest a new direction for semiconductor spintronics 
research.  

In the following paragraphs we argue that in high-mobility two-dimensional electron systems (2DES)
that have substantial Rashba \cite{rashbaham} spin-orbit coupling, spin currents always  
accompany charge currents. 
The Hamiltonian of a 2DES with Rashba spin-orbit coupling is given by \cite{rashbaham}
\begin{equation}
H=\frac{p^2}{2m}-\frac{\lambda}{\hbar}\vec{\sigma}\cdot(\hat{z}\times\vec{p}),
\label{rashham}
\end{equation}
where  $\lambda$ is the Rashba coupling constant, $\vec{\sigma}$ the Pauli matrices, $m$ the electron effective mass, and $\hat{z}$ the unit vector perpendicular to the 2DES plane.
The Rashba coupling strength in a 2DES can be modified by as much as 
50\% by a gate field \cite{nitta}. Recent observations of a  
spin-galvanic effect \cite{ganichev,ganichev2} and a spin-orbit coupling induced metal-insulator
transition in these systems \cite{koga} illustrate the potential importance of this
tunable interaction in semiconductor spintronics \cite{inoue}. 
The spin current we discuss is polarized in the direction perpendicular
to the two-dimensional plane and flows in the planar direction that is perpendicular to the charge 
current direction. It is therefore a spin Hall effect, but unlike the effect 
conceived by Hirsch \cite{hirsch}, it is purely intrinsic and does not rely on 
anisotropic scattering by impurities.  Remarkably, in the usual case when both
spin-orbit split Rashba bands are occupied, the spin-Hall conductivity has a 
universal value independent of both the 2DES density and the  Rashba coupling strength.  

\begin{figure}
\includegraphics[width=3.0in]{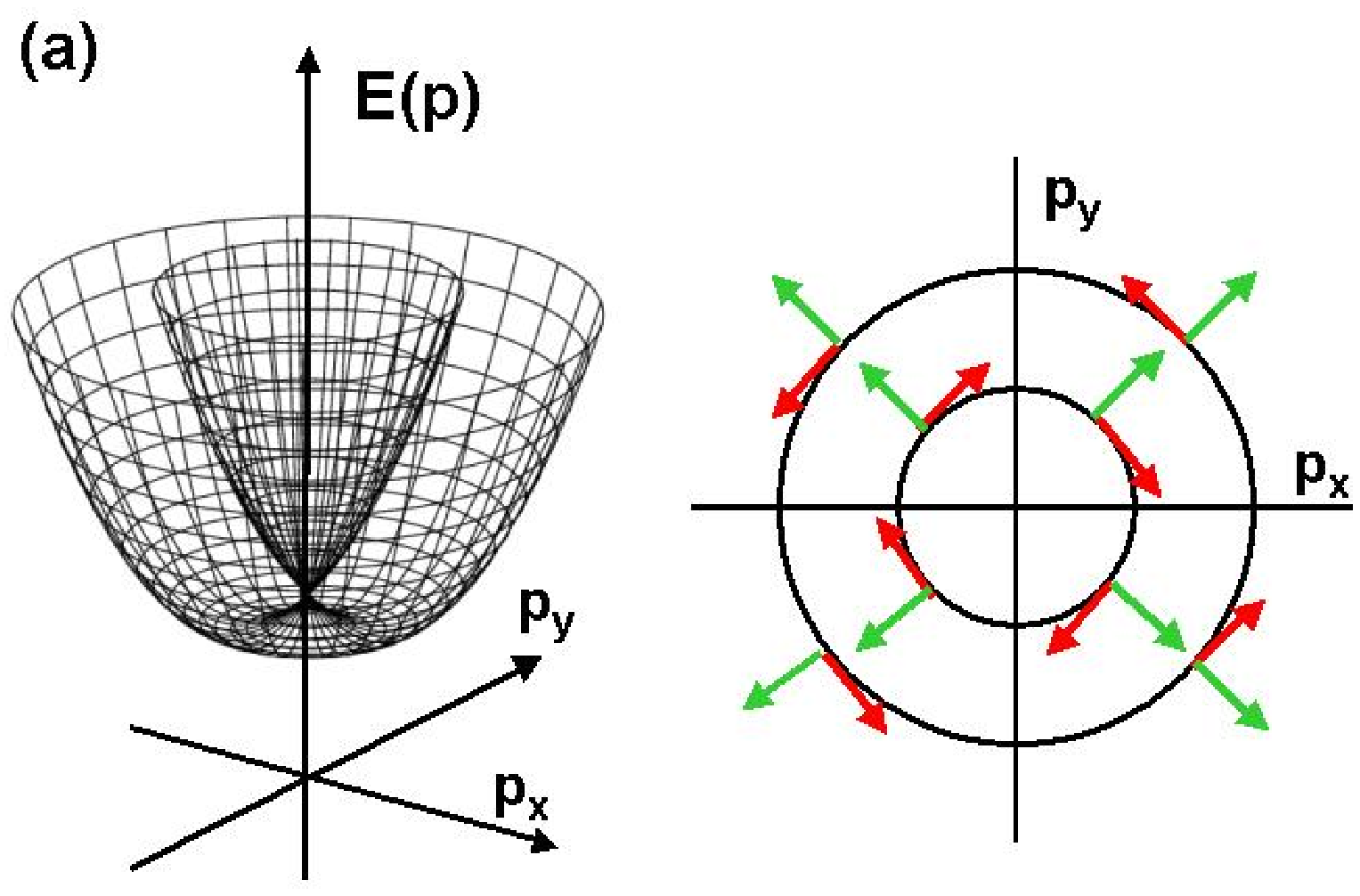}
\includegraphics[width=3.0in]{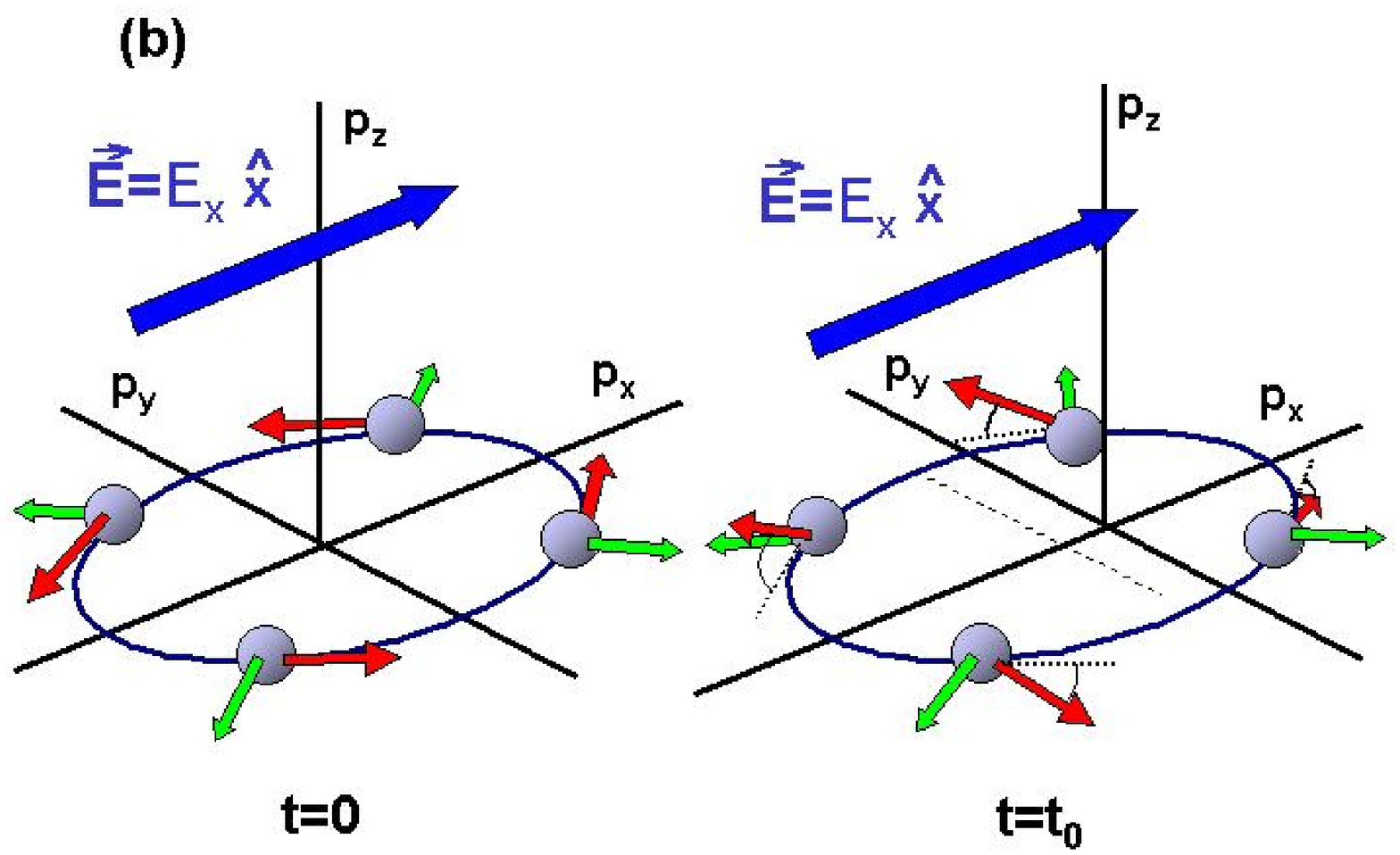}
\caption{
a) The 2D electronic eigenstates in a Rashba spin-orbit coupled system are labeled by 
momentum (green arrows).  For each momentum the two eigenspinors point in the azimuthal
direction (red arrows). b) In the presence
of an electric field (blue) the Fermi surface (blue circle) is displaced an amount $|eE_x t_0/\hbar|$ 
at time $t_0$ (shorter than typical scattering times).  While moving in momentum space, 
electrons experience an effective torque which tilts the spins up for
$p_y>0$ and down for $p_y<0$, creating a spin-current in the y-direction.}
\label{fig:one}
\end{figure}

The basic physics of this effect is illustrated schematically in 
Fig.~\ref{fig:one}.  In a translationally invariant 2DES, electronic eigenstates have
definite momentum and, because of spin-orbit coupling, a momentum dependent effective 
magnetic field that causes the spins (red arrows) to align perpendicular to the
momenta (green arrows), as illustrated in Fig.~\ref{fig:one} (a).
In the presence of an electric field, which we take to be in the $\hat x$ direction and indicate 
by blue arrows in Fig.~\ref{fig:one} (b), electrons are accelerated and drift 
through momentum space at the
rate $\dot{\vec p} = - e E \hat x$.  Our spin-Hall effect arises from the time dependence of the 
effective magnetic field experienced by the spin because of its motion in momentum space.  
For the Rashba Hamiltonian case of interest here, the effect can be understood most simply
by considering the Bloch equation of a spin-$1/2$ particle, as we explain in the following paragraph.  
More generally the effect arises from non-resonant interband contributions to the Kubo
formula expression for the spin Hall conductivity that survive in the static limit.
   
The dynamics of an electron spin in the presence of time-dependent 
Zeeman coupling is described by the Bloch equation: 
\begin{equation}
\frac{\hbar d \hat{n}}{dt} = \hat{n} \times \vec{\Delta}(t) + \alpha \; 
\frac{\hbar d \hat{n}}{dt} \times \hat{n} 
\label{lleq}
\end{equation}
where $\hat n$ is direction of the spin and $\alpha$ is a damping parameter that we assume is small. 
For the application we have in mind the $\vec{p}$ dependent 
Zeeman coupling term in the spin-Hamiltonian is $- \vec{s} \cdot \vec{\Delta}/\hbar$, where
$\vec{\Delta}=2\lambda/\hbar(\hat{z}\times\vec{p})$. 
For a Rashba effective magnetic field with magnitude $\Delta_1$ that initially points in the
$\hat x_1$ direction then tilts (arbitrarily slowly) slightly toward $\hat x_2$, where 
$\hat x_1$ and $\hat x_2$ are orthogonal in-plane directions, it follows from the linear 
response limit of Eq.(~\ref{lleq}) that 
\begin{eqnarray}
\frac{\hbar d n_2}{dt} &=& n_z \Delta_1 + \alpha \; dn_z/dt \nonumber \\
\frac{\hbar d n_z}{dt} &=& - \Delta_1 n_2 -\alpha \; dn_2/dt +\Delta_2,
\label{linearizedlleq}
\end{eqnarray}
where $\Delta_2=\vec{\Delta}\cdot\hat{x}_2$.   Solving these inhomogeneous coupled 
equations using a Greens function technique, it follows that 
to leading order in the slow-time dependences $n_2(t) = \Delta_2(t)/\Delta_1$, {\it i.e.}, the
$\hat x_2$-component of the spin rotates to follow the direction of the field, and that 
\begin{equation}
n_z(t) = \frac{1}{\Delta_1^2} \frac{\hbar d \Delta_2}{dt}.
\label{tilt1}
\end{equation}

Our intrinsic spin-Hall effect follows from Eq.(~\ref{tilt1}). 
When a Bloch electron moves through momentum space, its spin orientation
changes to follow the momentum-dependent effective field and also 
acquires a momentum-dependent $\hat{z}$-component.
We now show that in the case of Rashba spin-orbit coupling, this effect 
leads to an intrinsic spin-Hall conductivity that has a  
universal value in the limit of zero quasiparticle spectral broadening. 

For a given momentum $\vec{p}$, the spinor originally points in the azimuthal direction.
An electric field in the $\hat{x}$ direction ($\dot{p}_x =-eE_x$) changes the $y$-component 
of the $\vec{p}$-dependent effective field. 
Applying the adiabatic spin dynamics expressions explained above, identifying
the azimuthal direction in momentum space with $\hat x_1$ and the radial
direction with $\hat x_2$ 
we find that the $z$-component of the spin direction
for an electron in a state with momentum $\vec{p}$ is 
\begin{equation}
n_{z,\vec{p}} = \frac{-e \hbar^2 p_y E_x}{ 2\lambda p^3}.
\label{tilt}
\end{equation}
(Linear response theory applies for 
$ e E_x r_s \ll \Delta_1$ where $r_s$ is the interparticle spacing in the 2DES.) 
Summing over all occupied states the linear response of the $\hat{z}$ spin-polarization component 
vanishes because of the odd dependence of $n_{z}$ on $p_y$, as illustrated in Fig.~\ref{fig:one}, 
but the spin-current in the $\hat{y}$ direction is finite.

The Rashba Hamiltonian has two eigenstates for each momentum with eigenvalues $E_{\pm}={p}^2/2m \mp \Delta_1/2$;
the discussion above applies for the lower energy (labeled + for majority spin Rashba band) 
eigenstate while the higher energy (labeled -) eigenstate has the
opposite value of $n_{z,\vec{p}}$.  Since $\Delta_1$ is normally much smaller than the Fermi energy \cite{nitta},
only the annulus of momentum space that is occupied by just the lower energy band contributes to the spin-current.
In this case we find that the spin-current in the $\hat{y}$ direction is\cite{footnotevelocity}  
\begin{eqnarray}
j_{s,y} &=& \int_{annulus} \frac{d^2 \vec{p}}{(2 \pi \hbar)^2} \frac{\hbar n_{z,\vec{p}}}{2} \frac{p_y}{m} \nonumber \\
&=& \frac{-e E_x }{16 \pi \lambda m} (p_{F+}-p_{F-}),
\label{spincurrent}
\end{eqnarray}
where $p_{F+}$ and $p_{F-}$ are the Fermi momenta of the majority and minority spin
Rashba bands. We find that when both bands are occupied, i.e. when $n_{2D}>m^2\lambda^2/\pi\hbar^4
\equiv n_{2D}^*$,
$p_{F+}-p_{F-} = 2 m \lambda/\hbar$ 
and then the spin-Hall conductivity is  
\begin{equation}
\sigma_{sH} \equiv -\frac{j_{s,y}}{E_x} = \frac{e}{8 \pi}
\label{spinhall1}
\end{equation}
independent of both the Rashba coupling strength and of the 2DES density.  
For $n_{2D}<n_{2D}^*$ the upper Rashba band is depopulated. In this limit 
$p_{F-}$ and $p_{F+}$ are the interior and
exterior Fermi radii of the lowest Rashba split band, and $\sigma_{sH}$ 
vanishes linearly with the 2DES density:
\begin{eqnarray}
\sigma_{sH}=
\frac{e}{8\pi}\frac{n_{2D}}{n_{2D}^*}
\label{spinhall}
\end{eqnarray}

\begin{figure}
\includegraphics[width=3.4in]{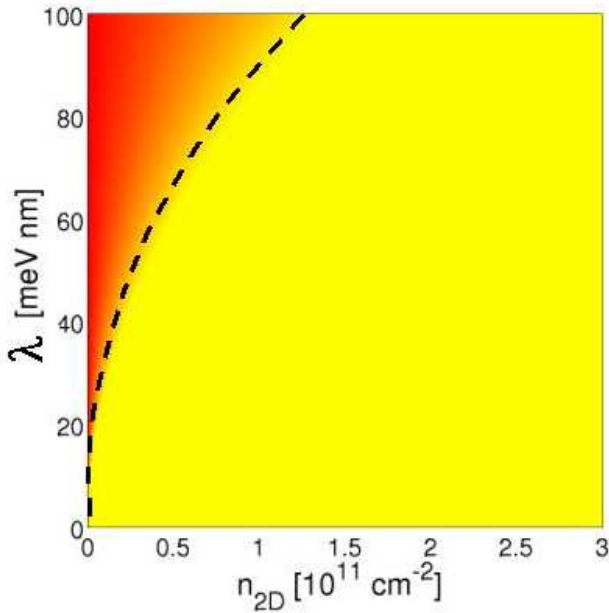}
\caption{Color plot of $\sigma_{sH}$ vs. the two-dimensional electron density $n_{2D}$ and
the Rashba coupling constant $\lambda$. The color scale is yellow =$e/8\pi$ and dark red=0.
The dash lines indicates the boundary between the universal region (constant $\sigma_{sH}$)
and the linear density dependence region.
Most of the current experimental samples are deep in the universal value region.
We expect the intrinsic spin-Hall contribution to be present only when the 
Rashba splitting is larger than the disorder broadening of the quasiparticle 
energy levels, {\it i.e.} when $ \lambda p_F \tau /\hbar^2 > 1$.}
\label{fig:two}
\end{figure}

The intrinsic spin-Hall conductivity of the Rashba model can also 
be evaluated by using transport theories that are valid for systems
with multiple spin-orbit spit bands, 
either the linear-response-theory Kubo-formula approach\cite{otherlongpaper}, 
or a generalized Boltzmann equation approach that accounts for 
anomalous contributions to wave-packet dynamics\cite{culcer2}. 
In the Kubo formalism approach, our universal intrinsic spin-Hall effect
comes from the static $\omega=0$ limit of the 
non-dissipative reactive term in the expression for the spin-current response 
to an electric field\cite{kuboref}:  
\begin{eqnarray}
\sigma^{\rm sH}_{xy}(\omega)&=&\frac{e\hbar}{V}\sum_{{\bf k},n\ne n'}
(f_{n',k}-f_{n,k})\nonumber\\&\times&
\frac{{\rm Im}[\langle n' k|
\hat{j}^z_{\rm spin\,\,x}|nk\rangle\langle nk| v_y|n' k\rangle]} 
{(E_{nk}-E_{n'k})(E_{nk}-E_{n'k}-\hbar \omega -i \eta)}
\label{SH}
\end{eqnarray}
where $n,n'$ are band indices, $\vec{j}^z_{\rm spin}=\frac{\hbar}{4}\{\sigma_z,\vec{v}\}$ is
the spin-current operator, $\omega$ and $\eta$ are set to zero in the dc clean limit, 
and the velocity operators at each $\vec{p}$ are given\cite{molenkamp} by 
\begin{eqnarray} 
\hbar v_x &=& \hbar {\partial{H(\vec{p})}/{\partial p_x}} = \hbar p_x/m - \lambda \sigma_y \nonumber \\
\hbar v_y &=& \hbar {\partial{H(\vec{p})}/{\partial p_y}} = \hbar p_y/m + \lambda \sigma_x 
\label{velocities}
\end{eqnarray} 
For the Rashba model $n,n' =\pm$ and the 
eigenspinors are spin coherent states given explicitly by
\begin{eqnarray}
|\mp,p\rangle&=&\frac{1}{\sqrt{2}}\left(
\begin{array}{c}
\pm ie^{-i\phi}\\1
\end{array}
\right)
\label{spinors}
\end{eqnarray}
where $\phi=\arctan {p_x/p_y}$.  The energy denominators in Eq.(~\ref{SH}) in this 
case are given by the Rashba splitting, and the velocity matrix elements can
be evaluated using Eq.(~\ref{velocities}) and Eq.(~\ref{spinors}). 
The integral over momentum-space can be evaluated easily because the energy 
denominators are independent of orientation and the intrinsic spin-Hall conductivity expressions
in Eqs. \ref{spinhall1} and \ref{spinhall} are recovered. 
This principal result of the Letter is summarized in Fig.~\ref{fig:two}
where $\sigma_{sH}$ is plotted as a function of carrier
density and Rashba coupling strength in the zero quasiparticle spectral broadening limit.
The Kubo formula analysis also makes it clear that, unlike the universal Hall conductivity
value on a 2DES quantum Hall plateau, the universality of the intrinsic spin Hall effect 
is not robust against disorder and will be reduced whenever the disorder broadening
is larger than the spin-orbit coupling splitting \cite{sinova,nikolai,john}.  

The intrinsic character of our spin Hall effect, compared to the extrinsic
character of the effect discussed originally by Hirsch \cite{hirsch}, is
analogous to the intrinsic character that we have recently proposed
for the anomalous Hall effect in some ferromagnets and strongly polarized paramagnets
\cite{sundaram,jungwirth,jungwirth2,sinova,nagaosa1,nagaosa2,culcer,luttinger}.  
In both cases the skew scattering contributions to 
the Hall conductivities can become important \cite{bruno,dietlahe}, when the overall electron
scattering rate is small and the steady state distribution function of the
current-carrying state is strongly disturbed compared to the equilibrium one.  
In the Kubo-formula approach these contributions to the spin-Hall 
conductivity appear as dissipative contributions from disorder scattering 
of Fermi-energy quasiparticles. 
A general and quantitative analysis of the disorder-potential-dependent 
interplay between intrinsic and skew-scattering contributions to anomalous Hall
and spin-Hall effects is a subject of a current research that is beyond the scope of this Letter. 

Several schemes have been proposed for measuring the spin-Hall effect
in metals \cite{hirsch,zhang,pareek} and these can be generalized to the semiconductor case.
In semiconductors the close relationship between optical properties\cite{jonker,molenkamp} and 
spin-polarizations opens up new possibilities for detecting non-equilibrium 
spins accumulated near contacts or near the sample perimeter by spin currents. 
Spatially resolved Faraday or Kerr effects
should be able to detect spin accumulations induced by the spin-currents we have evaluated.
As in the case of the ferromagnetic semiconductor anomalous Hall effect, the
origin of the intrinsic spin
Hall type effect is strong spin-orbit coupling.  
A sizable intrinsic spin Hall effect will occur in any paramagnetic material with strong 
spin-orbit coupling, including hole-doped bulk semiconductors \cite{murakami},
although the universal value we obtain here is a unique property of Rashba systems.

\begin{acknowledgments}
The authors would like to thank S. Murakami, N. Nagaosa, and S.-C. Zhang for 
sharing their results with us prior to publication. We also would like
to acknowledge insightful interactions with M.M. Trudys 
and G. T. B. Sapphire.  This work was supported by the Welch Foundation,
by the DOE under grant DE-FG03-02ER45958, and by the Grant Agency of the Czech Republic
under grant 202/02/0912, NSF under grant DMR0072115,  DOE No. DE-FG03-96ER45598,
and the Telecommunication and Information Task Force at TAMU.
\end{acknowledgments}


\begin{thebibliography}{37}
\bibitem{footnotevelocity}  The electron group velocity includes a term proportional to spin (see Eq. \ref{velocities})
which is dropped because in Eq. \ref{spincurrent} it is the spin-current which is included. This spin current
is formed by the anticommutator of the spin and the group velocity and the second term hence vanishes.
\end{thebibliography}

\end{document}